\documentclass[preprint,12pt]{elsarticle}

\usepackage{amssymb}
\usepackage{amsmath}
\usepackage{graphicx}
\usepackage{verbatim,color,amsfonts,latexsym}
\journal{Physica A}

\begin{document}

\begin{frontmatter}

%% Title, authors and addresses

%% use the tnoteref command within \title for footnotes;
%% use the tnotetext command for theassociated footnote;
%% use the fnref command within \author or \address for footnotes;
%% use the fntext command for theassociated footnote;
%% use the corref command within \author for corresponding author footnotes;
%% use the cortext command for theassociated footnote;
%% use the ead command for the email address,
%% and the form \ead[url] for the home page:
%% \title{Title\tnoteref{label1}}
%% \tnotetext[label1]{}
\author[label1]{C. N. Angstmann}
\ead{c.angstmann@unsw.edu.au}
\author[label1]{B. I. Henry \corref{cor}}
\ead{b.henry@unsw.edu.au}
\author[label1]{A. V. McGann}
\ead{a.mcgann@unsw.edu.au}
\address[label1]{UNSW Australia, Sydney, NSW 2052, Australia}

\cortext[cor]{Corresponding Author}
%\author[label1,label2,label3]{C. N. Angstmann\fnref{label1}, B. I. Henry\corref{cor2}, A. V. McGann\fnref{label2}}

%% \ead[url]{home page}
%% \fntext[label2]{}
%% \cortext[cor1]{}
%% \address{Address\fnref{label3}}
%% \fntext[label3]{}

\title{A Fractional-Order Infectivity SIR Model}

%% use optional labels to link authors explicitly to addresses:
%% \author[label1,label2]{}
%% \address[label1]{}
%% \address[label2]{}

%\author{C. N. Angstmann, B. I. Henry, A. V. McGann}

\begin{abstract}
Fractional-order SIR models have become increasingly popular in the literature in recent years, however unlike the standard SIR model, they often lack a derivation from an underlying stochastic process. Here we derive a fractional-order infectivity SIR model from a stochastic process that incorporates a time-since-infection dependence on the infectivity of individuals. The fractional derivative appears in the generalised master equations of a continuous time random walk through SIR compartments, with a power-law function in the infectivity. %The equations also allow for an arbitrary waiting time from becoming infected to being removed.
We show that this model can also be formulated as an infection-age structured Kermack-McKendrick integro-differential SIR model. Under the appropriate limit the fractional infectivity model reduces to the standard ordinary differential equation SIR model.
\end{abstract}


\end{frontmatter}

%% \linenumbers

%% main text
\setcounter{equation}{0}
\section{Introduction}

The SIR model was first introduced by Kermack and McKendrick in 1927 \cite{KM1927} as a mathematical model of an epidemic. This model is the cornerstone of mathematical epidemiology with many variations developed \cite{H2000,DH2000}. In their later work, Kermack and McKendrick incorporated an age-structure into the model through the use of integro-differential equations \cite{KM1932,KM1933}. As a special case, this allows for the model to account for diseases in which the rate of recovery is dependent on how long an individual has been infected. In previous work we showed that if the time infected is power law distributed then the governing system of equations can be written as a coupled set of fractional order differential equations \cite{AHM2015}. Here we consider an alternate disease process accounting for infectivity as a function of the time since infection
which also leads to the inclusion of fractional derivatives in the governing equations.

In the classic SIR model with vital dynamics, the population is broken up into three compartments, Susceptible (S),  Infectious (I), and Recovered (R) \cite{KM1927}. The population is born into the S compartment, moving into the I compartment after infection, and into the R compartment upon recovery. Individuals may be removed from any compartment through death. The dynamics of this model can be expressed as a set of coupled ordinary differential equations (ODEs). From a stochastic process perspective, an individual's transition through the compartments can be treated as a directed generalised continuous time random walk (CTRW) \cite{AHM2015}. In this CTRW each individual will wait for a random time before transitioning to the next compartment. This is an extension of the classic CTRW in which particles perform an unbiased walk on a lattice with stochastic waiting times between steps \cite{MW1965,SL1973}. The limit process of a CTRW with power-law waiting times can result in fractional diffusion equations \cite{MK2000} and fractional reaction-diffusion equations \cite{SSS2006,HLW2006,F2010,AYL2010,ADH2013mmnp}. 

In some disease processes \cite{R2013} the dynamics of the system is dependent on both the current state and history of the system. The classic SIR model is insufficient for dealing with such disease processes. The age-structured approach of Kermack and McKendrick can be used, at the expense of an additional time dimension for the problem. A more recent approach has been to incorporate a history dependence into the dynamics by generalising the classic coupled ODEs using fractional time derivatives \cite{AR2012,AE2013,DU2011,D2013,GG2014,GM2014,ZZ2013}. This generalisation is typically achieved by replacing the integer order time derivative with a fractional order Caputo derivative \cite{P1999}. The Caputo derivative of a function is dependent on the entire history of the function and as such the generalised models will include a history dependence in the dynamics. While these fractional models are of interest mathematically, and numerous methods have been developed for the solution of fractional differential equations \cite{KST2006}, the modelling approach may not match the underlying physical process. 
Another issue that needs to be addressed in modelling disease processes and other coupled systems with fractional order differential equations is that of matching dimensions of parameters with fractional rates \cite{D2013,SRC2015}. 

A more sophisticated approach to provide fractional calculus compartment models \cite{DM2009,DMM2010} is to
replace constant parameters in standard ODE  models with time dependent parameters, then to integrate the 
equations, introducing additional kernels in the integrand that depend on two time parameters. Power law  kernels
typically result in fractional integrals and then after taking time derivatives on both sides of the equation, coupled fractional order differential equations are obtained. This approach, which corrects the dimension problem, was recently used to provide a fractional order model for the spread of dengue fever \cite{SRC2015}. Some of the fractional calculus compartment models derived in this way have
been shown to yield unbounded concentrations and  input functions are needed to compensate for this \cite{HH2013}.

In this work we derive an SIR model from a stochastic process, showing how a fractional-order derivative can be incorporated into the infectivity rate of the disease. This is an extension of our recent work in which we incorporated a fractional derivative into the recovery rate in an SIR model to incorporate the effects of chronic infection \cite{AHM2015}. In section 2 we derive a general infectivity SIR model from a CTRW and show the consistency of the derived model with a Kermack-McKendrick age-structured SIR model. A power-law rate is considered for the infectivity in section 3. This results in the inclusion of a fractional-order derivative in the infectivity term of the model. The equilibrium states of the system are found in section 4.

\vspace*{0.5cm}
\setcounter{equation}{0}
\section{Derivation}
\label{sec:derivation}
In order to incorporate the fractional-order infectivity, we first derive master equations for an SIR model with a general infectivity rate, using a stochastic process. We consider a generalised continuous time random walk where an individual transitions through the three compartments, waiting a random time in each compartment. The model considers an ensemble of such individuals. An individual who has been infectious since time $t'$ will infect a particular susceptible person in the time interval $t$ to $t+\delta t$, with probability $\sigma(t,t')\delta t+ o(\delta t)$. The transmission rate per infected individual, $\sigma(t,t')$ is dependent on both the time of infection, $t'$, and current time, $t$. Given that there are $S(t)$ susceptible people at time $t$ then in the time interval $t$ to $t+\delta t$ the expected number of new infections per infected individual will be $\sigma(t,t') S(t) \delta t+o(\delta t)$.

The number of individuals entering the infected compartment at time $t$, i.e. the flux into I, will be represented as $q^+(I,t)$. This can be recursively constructed from the flux at earlier times by, 
\begin{equation}
\label{eq_flux}
q^{+}(I,t)=\int_{-\infty}^{t}\sigma(t,t') S(t)\Phi(t,t')q^{+}(I,t')dt',
\end{equation}
where $\Phi(t,t')$ is the survival function that an individual infected at a prior time $t'$ remains infected at time $t$. Considering the initial distribution of infected individuals in the population, we 
let $i(-t',0)$ be the number of individuals who became infected at time $t'<0$ and who are still infected at time $0$, hence,
\begin{equation}
q^+(I,t')= \frac{i(-t',0)}{\Phi(0,t')}, \hspace{10pt}t'<0. 
\end{equation}
Hence we can split Eq. (\ref{eq_flux}) into, 
\begin{equation}
\label{eq_flux2}
q^{+}(I,t)=\int_{0}^{t}\sigma(t,t') S(t)\Phi(t,t')q^{+}(I,t')dt'+\int_{-\infty}^{0}\sigma(t,t') S(t)\frac{\Phi(t,t')}{\Phi(0,t')}i(-t',0)dt'.
\end{equation}
It is natural to assume that the rate of infection, $\sigma(t,t')$, is dependent on the time $t$ to account for environmental changes in time. It is expected that $\sigma(t,t')$ may also depend on the age of infection, $t-t'$, to account for the natural course of the disease. In the following we incorporate these effects by writing,
\begin{equation}
\label{eq_infec}
\sigma(t,t')=\omega(t)\rho(t-t').
\end{equation}
Noting that an individual may leave the infected compartment in two ways, either they die or they recover from the disease and assuming these effects are independent we can write the survival function as,
\begin{equation}
\label{eq_PHI}
\Phi(t,t')=\phi(t,t')\theta(t,t').
\end{equation}
Here $\phi(t,t')$ is the probability of surviving the transition to the $R$ compartment from time $t'$ to time $t$, and $\theta(t,t')$ is the probability of surviving the death transition from time $t'$ until time $t$.
We will assume that the recovery and death survival take the form,
\begin{align}
\label{eq_theta}
\theta(t,t')&=e^{-\int_{t'}^{t}\gamma(s)ds},\\
\label{eq_phi}
\phi(t,t')&=e^{-\int_{t'}^{t}\mu(s)ds}.
\end{align}
From this it follows that $\Phi(t,t')$ satisfies the semi-group property,
\begin{equation}
\label{eq_semig}
\Phi(t,t')=\Phi(t,s)\Phi(s,t'),\;\;\forall\; t'<s<t.
\end{equation}

For an individual to be infected at time $t$ they must have become infected at some time prior to $t$ and not yet transitioned into the removed compartment nor died. Hence the number of individuals in the $I$ compartment at time $t$ can be expressed recursively using the flux as,
\begin{equation}
\label{eq_totI}
I(t)=I_0(t)+\int_{0}^{t}\Phi(t,t')q^{+}(I,t')dt'.
\end{equation}
%This sentence is very vague
In which $I_0(t)$ is the number of initially infected individuals, $i(-t',0)$, whose infection has persisted until time $t$, expressed as,
\begin{equation}
\begin{split}
I_0(t)&=\int_{-\infty}^{0}\frac{\Phi(t,t')}{\Phi(0,t')}i(-t',0)dt',\\
&=\Phi(t,0)\int_{-\infty}^{0}i(-t',0)dt'.
\end{split}
\end{equation}
In order to recover the master equations governing the dynamics, 
we differentiate Eq. (\ref{eq_totI}) to produce,
\begin{equation}
\label{eq_didt}
\begin{split}
\frac{dI(t)}{dt}&=q^{+}(I,t)-\left(\mu(t)+\gamma(t)\right)\int_{0}^{t}\phi(t,t')\theta(t,t')q^{+}(I,t')dt'-(\gamma(t)+\mu(t))I_0(t), \\
&=q^{+}(I,t)-(\mu(t)+\gamma(t))I(t).
\end{split}
\end{equation}
Substituting Eqs. (\ref{eq_flux2}), \eqref{eq_infec} into Eq. (\ref{eq_didt}) gives,
\begin{equation}
\label{eq_didt2}
\begin{split}
\frac{dI(t)}{dt}=&\int_{0}^{t}\omega(t) \rho(t-t') S(t)\Phi(t,t')q^{+}(I,t')dt'+\int_{-\infty}^{0}\omega(t)\rho(t-t')S(t)\Phi(t,0)i(-t',0)dt'\\&-(\mu(t)+\gamma(t))I(t).
\end{split}
\end{equation}
In order to obtain a generalised master equation we need to express the right hand side of this equation in terms of $I(t)$. 
We can write Eq. (\ref{eq_totI}) using Eq. (\ref{eq_semig}) as,
\begin{equation}
\frac{I(t)}{\Phi(t,0)}=\frac{I_0(t)}{\Phi(t,0)}+\int_{0}^{t}\frac{q^{+}(I,t')}{\Phi(t',0)}dt'.
\end{equation}
Taking the Laplace transform from $t$ to $s$ then gives, 
\begin{equation}
\label{eq_ltq}
\mathcal{L}\left\{\frac{q^{+}(I,t)}{\Phi(t,0)}\right\}=s\mathcal{L}\left\{\frac{I(t)-I_0(t)}{\Phi(t,0)}\right\}.
\end{equation}
Returning to the first integral of Eq. \eqref{eq_didt2} we can rewrite it using Laplace transforms as,
\begin{equation}
\label{eq_Ikap}
\omega(t)S(t)\int_{0}^{t}\rho(t-t')\frac{q^{+}(I,t')}{\Phi(t',0)}dt'=\omega(t)S(t)\mathcal{L}^{-1}\left\{\mathcal{L}\{\rho(t)\}\mathcal{L}\left\{\frac{q^{+}(I,t)}{\Phi(t,0)}\right\}\right\}.
\end{equation}
Making use of Eq. \eqref{eq_ltq} this becomes,
\begin{equation}
\begin{split}
\label{eq_kernd}
\omega(t)S(t)\mathcal{L}^{-1}\left\{\mathcal{L}\{\rho(t)\}\mathcal{L}\left\{\frac{q^{+}(I,t)}{\Phi(t,0)}\right\}\right\} &= \omega(t)S(t)\mathcal{L}^{-1}\left\{s\mathcal{L}\{\rho(t)\}\mathcal{L}\left\{\frac{I(t)-I_0(t)}{\Phi(t,0)}\right\}\right\},\\
&=\omega(t)S(t)\int_{0}^{t}\kappa(t-t')\frac{I(t')-I_0(t')}{\Phi(t',0)}dt',
\end{split}
\end{equation}
where we have defined, 
\begin{equation}
\label{eq_kern}
\kappa(t)=\mathcal{L}^{-1}\{s\mathcal{L}\{\rho(t)\}\}.
\end{equation}
Using Eq. (\ref{eq_kernd}) and Eq. \eqref{eq_kern}, in Eq. (\ref{eq_didt2}), we obtain the master equation,
\begin{equation}
\begin{split}
\frac{dI(t)}{dt}=&\omega(t) S(t) \Phi(t,0)\left(\int_{0}^{t}\kappa(t-t')\frac{I(t')-I_0(t')}{\Phi(t',0)}dt'+ \int_{-\infty}^{0}\rho(t-t')i(-t',0)dt'\right)\\&-\mu(t)I(t)-\gamma(t)I(t).
\end{split}
\end{equation}
Noting that $\frac{I_0(t)}{\Phi(t,0)}$ is a constant and using Eq. (\ref{eq_kern}) this may be written as,
\begin{equation}
\label{eq_gme}
\begin{split}
\frac{dI(t)}{dt}=&\omega(t) S(t) \Phi(t,0)\left(\int_{0}^{t}\kappa(t-t')\frac{I(t')}{\Phi(t',0)}dt'+ \int_{-\infty}^{0}\left(\rho(t-t')-\rho(t)\right)i(-t',0)dt'\right)\\&-\mu(t)I(t)-\gamma(t)I(t).
\end{split}
\end{equation}
This equation is the generalised master equation that describes the time evolution of the number of infected individuals in an SIR model with arbitrary time dependent infectivity and recovery. 
As individuals may only enter the infected compartment from the susceptible compartment, there must be a corresponding decrease in the number of individuals in the susceptible compartment. Accounting for vital dynamics, the differential equation for the susceptible population is then given by,
\begin{equation}
\label{eq_genS}
\begin{split}
\frac{dS(t)}{dt}=&\lambda(t)-\omega(t) S(t) \Phi(t,0)\left(\int_{0}^{t}\kappa(t-t')\frac{I(t')}{\Phi(t',0)}dt'+\int_{-\infty}^{0}(\rho(t-t')-\rho(t))i(-t',0)dt'\right)\\&-\gamma(t)S(t),
\end{split}
\end{equation}
where $\lambda(t)\geq 0$ is the birth rate and $\gamma(t) \geq 0$ is the per capita death rate.
Using a similar balance between the infected and recovered compartment, the differential equation for the recovered compartment is,
\begin{equation}
\label{eq_genR}
\frac{dR(t)}{dt}=\mu(t)I(t)-\gamma(t)R(t).
\end{equation}
\\Taking the initial condition $i(-t,0)=i_0 \delta(-t)$, where $\delta(-t)$ is a Dirac delta function and $i_0$ is a constant, these equations further simplify to give,
\begin{align}
\label{eq_genS_delta}
\frac{dS(t)}{dt}=&\lambda(t)-\omega(t) S(t) \Phi(t,0)\left(\int_{0}^{t}\kappa(t-t')\frac{I(t')}{\Phi(t',0)}dt'\right)-\gamma(t)S(t),\\
\label{eq_genI_delta}
\frac{dI(t)}{dt}=&\omega(t) S(t) \Phi(t,0)\left(\int_{0}^{t}\kappa(t-t')\frac{I(t')}{\Phi(t',0)}dt'\right)-\mu(t)I(t)-\gamma(t)I(t),\\
\label{eq_genR_delta}
\frac{dR(t)}{dt}=&\mu(t)I(t)-\gamma(t)R(t).
\end{align}

\vspace*{0.5cm}
\subsection{Structured SIR}
Here we show how the master equations, Eqs. \eqref{eq_genS_delta}, \eqref{eq_genI_delta}, \eqref{eq_genR_delta}, can be reduced to the Kermack and McKendrick age-structured SIR model \cite{KM1932} equations  given by, \begin{eqnarray}
\label{eq_KMS}
\frac{d S}{dt}&=&\lambda-S(t)\int_{0}^{\infty}\nu(t,a) i(a,t)da-\gamma S(t),\\
\label{eq_KMi}
\frac{\partial i}{\partial t}&+&\frac{\partial i}{\partial a}=-\beta(a) i(a,t)-\gamma i(a,t),\\
\label{eq_KMR}
\frac{dR}{dt}&=&\int_{0}^{\infty}\beta(a) i(a,t) da-\gamma R(t),\\
\label{eq_KMI}
I(t)&=&\int_{0}^{\infty}i(a,t)da.
\end{eqnarray}
In this model we consider $i(a,t)$ to be the number of the individuals infected at time $t$ who have been infected for length of time $a$. 
%We can recover the Eqs. (\ref{eq_gme}), (\ref{eq_genS}), (\ref{eq_genR}) by letting,
To show how Eq. (\ref{eq_genI_delta})  reduces to Eq. (\ref{eq_KMi}) we set $i(a,t)$ to,
\begin{align}
\label{eq_lili}
i(a,t)&=\Phi(t,t-a)q^+(I,t-a).
\end{align}
This allows us to see that $i(0,t)=q^+(I,t)$.  Integrating Eq. (\ref{eq_KMi}) with respect to $a$, using Eq. \eqref{eq_KMI} and equating $\beta(a)=\mu$ yields,
\begin{equation}
\label{eq_Ikerstruct}
\frac{dI(t)}{dt}=q^+(I,t)-\mu\int_0^{\infty}\Phi(t,t-a)q^+(I,t-a)da -\gamma I(t). 
\end{equation}
By taking a change in variable to $t'=t-a$ and making use of Eqs. \eqref{eq_flux2} and \eqref{eq_infec} we arrive at,
\begin{equation}
\begin{split}
\frac{dI(t)}{dt}=&\int_{0}^{t}\rho(t-t')\omega(t)S(t)\Phi(t,t')q^{+}(I,t')dt'+\int_{-\infty}^{0}\rho(t-t')\omega(t)S(t)\frac{\Phi(t,t')}{\Phi(0,t')}i(-t',0)dt'\\&-\mu\int_{-\infty}^{t}\Phi(t,t')q^+(I,t')dt' -\gamma I(t).
\end{split}
\end{equation}
We further simplify this expression by using Eqs. \eqref{eq_ltq}, \eqref{eq_Ikap} and \eqref{eq_kernd} and taking the initial condition to be $i(-t',0)=i_0 \delta(-t')$. This yields,
%Taking the initial condition to be $i(-t',0)=i_0 \delta(-t')$ simplifies the equation to,
\begin{equation}
\begin{split}
\frac{dI(t)}{dt}=&\omega(t)S(t)\int_{0}^{t}\kappa(t-t')\frac{I(t')}{\Phi(t',0)}dt'-\mu I(t) -\gamma I(t),
\end{split}
\end{equation}
which is a speical case of Eq. \eqref{eq_genI_delta}. To show how Eq. \eqref{eq_genS_delta} reduces to Eq. \eqref{eq_KMS} we consider a change of variable $a=t-t'$, hence we can rewrite Eq. \eqref{eq_KMS} as,
\begin{equation}
\frac{dS(t)}{dt}=\lambda-S(t)\int_{-\infty}^{t}\nu(t,t-t')\Phi(t,t')q^+(I,t')dt'-\gamma S(t),
\end{equation}
which is equivalent to Eq. \eqref{eq_genS_delta} if $\nu(t,t-t')=\sigma(t,t')$. Finally to recover Eq. \eqref{eq_genR_delta} from Eq. \eqref{eq_KMR} we make use of Eq. \eqref{eq_lili} and the change of variable $a=t-t'$, resulting in,
\begin{equation}
\frac{dR(t)}{dt}=\mu I(t)-\gamma R(t).
\end{equation}

\vspace*{0.5cm}
\setcounter{equation}{0}
\section{Fractional Infectivity SIR}
\label{sec:fract}
The general master equations given in Eqs. \eqref{eq_genS_delta}, \eqref{eq_genI_delta}, \eqref{eq_genR_delta} reduce to the classic SIR ODEs if $\rho(t)=\rho$, a constant. This can be seen from Eq. \eqref{eq_kern} where the corresponding memory kernel reduces to,
\begin{equation}
\kappa(t)= \rho \delta(t).
\end{equation}
If $\rho(t)$ is a power-law of the form,
\begin{equation}
\label{eq_ML_rho}
\rho(t)=\frac{t^{\alpha-1}}{\Gamma(\alpha)}, \quad 0 <\alpha \leq 1,
\end{equation} then the general master equations reduce to a set of fractional-order differential equations. The memory kernel following from Eq. \eqref{eq_kern} with power-law $\rho(t)$ given by Eq. \eqref{eq_kern} has Laplace transform,
\begin{equation}
\label{eq_memkerML}
\mathcal{L}_t\{\kappa(t)\}=s^{1-\alpha}.
\end{equation}
Hence the integral in Eqs. \eqref{eq_genS_delta}, \eqref{eq_genI_delta} can be written as follows,
\begin{align}
\int_{0}^{t}\kappa(t-t')\frac{I(t')}{\Phi(t',0)}dt' &= \int_0^t \kappa(t-t')\frac{I(t')}{\Phi(t,0)}dt',
\\&=\mathcal{L}_s^{-1}\left\{s^{1-\alpha}\mathcal{L}_t\left\{\frac{I(t')}{\Phi(t',0)}\right\}\right\}.
\end{align}
To evaluate the inverse Laplace transform in the above equation we use the result \cite{P1999},
\begin{equation}
\,_0\mathcal{D}_{t}^{1-\alpha}f(t)=\mathcal{L}^{-1}_s\{s^{1-\alpha}\mathcal{L}_t\{f(t)\}\}-\mathcal{L}^{-1}_s\{\,_0\mathcal{D}_t^{-\alpha}f(t)\big|_{t=0^+}\},
\end{equation}
where,
\begin{align}
\label{eq_fracdef}
\,_0\mathcal{D}_{t}^{1-\alpha}f(t)= \frac{1}{\Gamma(\alpha)}\dfrac{d}{dt}\int_0^t (t-t')^{\alpha-1} f(t')dt',
\end{align}
is  the Riemann-Liouville fractional derivative.
In the following we will assume that the fractional integral, $\,_0\mathcal{D}_t^{-\alpha}f(t)\big|_{t=0^+}$ is zero, in which case we can write,
\begin{align}
\label{eq_fracdef_mem}
\int_{0}^{t}\kappa(t-t')\frac{I(t')}{\Phi(t',0)}dt' =\,_0\mathcal{D}_{t}^{1-\alpha}\left(\frac{I(t')}{\Phi(t',0)}\right).
\end{align}
Substituting Eq. \eqref{eq_fracdef_mem} into the generalised master equations Eqs. \eqref{eq_genS_delta} and \eqref{eq_genI_delta} yields the fractional order infectivity SIR model, 
\begin{align}
\label{eq_fracS}
\frac{dS(t)}{dt}&=\lambda(t) -\omega(t)S(t)\Phi(t,0)\,_0\mathcal{D}_{t}^{1-\alpha}\left(\frac{I(t)}{\Phi(t,0)}\right) -\gamma(t) S(t),\\
\label{eq_fracI}
\frac{dI(t)}{dt} &=\omega(t)S(t)\Phi(t,0)\,_0\mathcal{D}_{t}^{1-\alpha}\left(\frac{I(t)}{\Phi(t,0)}\right)-\mu(t)I(t)-\gamma(t) I(t),\\
\label{eq_fracR}
\frac{dR(t)}{dt}&=\mu(t)I(t)-\gamma(t)R(t).
\end{align}
\subsection{Dimensionality}
An aspect of fractional SIR models that warrants further consideration is the dimensionality of the parameters. A time derivative of order one has dimension of [time]$^{-1}$, a fractional derivative of order $\alpha$, either a Caputo or Riemann-Liouville, will have a dimension of [time]$^{-\alpha}$. Hence the inclusion of fractional derivatives necessitates the redefinition of parameters in the associated models. This may lead to complications when considering the physical interpretation of rates.

In the fractional model derived above, we consider the equation for change in the number of infected individuals over time, Eq.  \eqref{eq_fracI}. As we have a order one time derivative of a population on the left hand side its dimension is [population][time]$^{-1}$. Thus the dimension of the right hand side must be the same. It is clear that the recovery and death rates, $\mu(t)$, and $\gamma(t)$,  must have dimension [time]$^{-1}$ as $I(t)$ has the dimension [population].

For the infectivity term, it is clear that the dimension of $S(t)$ is [population], and the fractional derivative $\,_0\mathcal{D}_{t}^{1-\alpha}\left(\frac{I(t)}{\Phi(t,0)}\right)$  is [population][time]$^{\alpha-1}$. In order for the dimensions of the infectivity term to be consistent with the model, we are left with $\omega(t)$ having dimension [population]$^{-1}$[time]$^{-\alpha}$. 
We note that the dimensions of  the infectivity rate per infected individual of the disease, $\sigma(t,t')=\omega(t)\rho(t-t')$, is [population]$^{-1}$[time]$^{-1}$, regardless of the fractional $\alpha$ exponent. 

\vspace*{0.5cm}
\setcounter{equation}{0}
\section{Equilibrium State Analysis} 
\label{sec:steady} 
The set of fractional infectivity SIR Eqs. \eqref{eq_fracS}, \eqref{eq_fracI}, \eqref{eq_fracR} are a non-autonomous dynamical system. This set up creates difficulty in finding the equilibrium states hence we will simplify the model by taking the birth, death, recovery and contact rates to be constant, i.e. $\lambda(t)=\lambda$, $\gamma(t)=\gamma$, $\mu(t)=\mu$ and $\omega(t)=\omega_\alpha$ respectively, where $\omega_\alpha$ represents the dependence of the chosen $\alpha$ exponent on $\omega(t)$, due to dimensionality considerations. Hence the model becomes,
\begin{align}
\label{eq_fracS_const}
\frac{dS(t)}{dt}&=\lambda-\omega_\alpha S(t)\Phi(t,0)\,_0\mathcal{D}_{t}^{1-\alpha}\left(\frac{I(t)}{\Phi(t,0)}\right) -\gamma S(t),\\
\label{eq_fracI_const}
\frac{dI(t)}{dt} &=\omega_\alpha S(t)\Phi(t,0)\,_0\mathcal{D}_{t}^{1-\alpha}\left(\frac{I(t)}{\Phi(t,0)}\right)-\mu I(t)-\gamma I(t),\\
\label{eq_fracR_const}
\frac{dR(t)}{dt}&=\mu I(t)-\gamma R(t).
\end{align}
This can be simplified further using Eqs. \eqref{eq_PHI}, \eqref{eq_theta}, \eqref{eq_phi} to rewrite,
\begin{equation}
\Phi(t,0)=e^{-(\gamma+\mu)t} .
\end{equation}
The equilibrium state, $(S^*,I^*,R^*)$, is defined by,
\begin{equation*}
\lim_{t \rightarrow \infty}S(t)=S^*,\hspace{25pt}\lim_{t \rightarrow \infty}I(t)=I^*,\hspace{25pt}\lim_{t \rightarrow \infty}R(t)=R^*.
\end{equation*}
Taking the limit as $t \rightarrow \infty$ of Eqs. \eqref{eq_fracS_const}, \eqref{eq_fracI_const} and  \eqref{eq_fracR_const} reduces the equations to,
\begin{align}
\label{eq_stS}
0&=\lambda-\lim_{t\rightarrow \infty}\omega_\alpha S(t)e^{-(\gamma+\mu)t}\,_0\mathcal{D}_{t}^{1-\alpha}\left(e^{(\gamma+\mu)t}I(t)\right)-\gamma S^*,\\
\label{eq_stI}
0&=\lim_{t\rightarrow \infty}\omega_\alpha S(t)e^{-(\gamma+\mu)t}\,_0\mathcal{D}_{t}^{1-\alpha}\left(e^{(\gamma+\mu)t}I(t)\right)-(\gamma+\mu)I^*,\\
\label{eq_stR}
0&=\mu I^*-\gamma R^*.
\end{align}
We are able to split the remaining limit into,
\begin{equation*}
\lim_{t\rightarrow \infty}\omega_\alpha S(t)e^{-\nu t}\,_0\mathcal{D}_{t}^{1-\alpha}\left(e^{\nu t}I(t)\right)=\left(\lim_{t\rightarrow \infty}\omega_\alpha S(t)\right)\left(\lim_{t\rightarrow \infty}e^{-\nu t}\,_0\mathcal{D}_{t}^{1-\alpha}\left(e^{\nu t}I(t)\right)\right),
\end{equation*}
where $\nu=\gamma+\mu$. Using the result of \cite{AHM2015},
\begin{equation*}
\label{eq_lim_lap}
\lim_{t\to\infty} e^{-\nu t}\,_0\mathcal{D}_{t}^{1-\alpha}\left(e^{\nu t} I(t)\right)=\nu^{1-\alpha}I^*,
\end{equation*}
and trivially we have $\lim_{t\rightarrow \infty}\omega_\alpha S(t)=\omega_\alpha S^*$, hence,
\begin{equation}
\label{eq_final_lim}
\lim_{t\rightarrow \infty}\omega_\alpha S(t)e^{-\nu t}\,_0\mathcal{D}_{t}^{1-\alpha}\left(e^{\nu t}I(t)\right)=\omega_\alpha (\gamma+\mu)^{1-\alpha} S^* I^*.
\end{equation}
Substituting Eq. \eqref{eq_final_lim} into Eqs. \eqref{eq_stS}, \eqref{eq_stI}, \eqref{eq_stR} yields,
\begin{align}
0&=\lambda-\omega_\alpha (\mu+\gamma)^{1-\alpha} S^* I^*-\gamma S^*,\\
0&=\omega_\alpha (\mu+\gamma)^{1-\alpha} S^* I^*-(\mu+\gamma)I^*,\\
0&=\mu I^* -\gamma R^*.
\end{align}
Solving these equations reveals two equilibrium states, the disease free state,
\begin{equation} \label{eq_ss1}
S^*=\frac{\lambda}{\gamma},\;\;\;I^*=0,\;\;\;R^*=0,
\end{equation}
and the endemic state,
\begin{equation}
\label{eq_ss2}
S^{*}=\frac{\mu+\gamma}{\omega_\alpha(\mu+\gamma)^{1-\alpha}},\;\;\;I^{*}=\frac{\lambda}{\mu+\gamma}-\frac{\gamma}{\omega_\alpha(\mu+\gamma)^{1-\alpha}},\;\;\;R^{*}=\frac{\mu}{\gamma}\left(\frac{\lambda}{\mu+\gamma}-\frac{\gamma}{\omega_\alpha(\mu+\gamma)^{1-\alpha}}\right).
\end{equation}
The disease free equilibrium state is non-negative for all valid system parameters. However, the endemic equilibrium is only non-negative if,\begin{equation}
\frac{\lambda \omega_\alpha}{\gamma} >(\mu+\gamma)^{\alpha}.
\end{equation}
In the case where $\alpha=1$ the equilibrium states reduce to the steady states of the standard SIR ODE model with vital dynamics.
We anticipate that, similar to the fractional recovery SIR model \cite{AHM2015}, the endemic equilibrium state will be an asymptotically stable state for all parameters where it is non-negative.
\vspace*{0.5cm}
\setcounter{equation}{0}

\section{Summary and Discussion}

In this paper, starting from a stochastic process, we have derived an SIR model where the evolution equations incorporate a fractional order derivative. 
This derivative, which appears in the flux into the infected compartment, arises from a power law dependence in the infectivity. 
We have shown that this fractional order infectivity SIR model can be written as an age structured SIR model. The dimensions
of the parameters in the fractional model depend on the order of the fractional derivative. The fractional model permits both a disease free, and an endemic, long time equilibrium state, dependent on the system parameters.
The assumptions that give rise to the fractional derivative could be experimentally validated from epidemiological studies by estimating the infectivity $\sigma(t,t')$ as a function of time $t$, and time of infection $t'$.

\vspace*{0.5cm}
\setcounter{equation}{0}
\section*{Acknowledgments}
This work was supported by the Australian Commonwealth Government (ARC No. DP130100595). 
\vspace*{0.5cm}
\setcounter{equation}{0}
%% The Appendices part is started with the command \appendix;
%% appendix sections are then done as normal sections
%% \appendix

%% \section{}
%% \label{}

%% If you have bibdatabase file and want bibtex to generate the
%% bibitems, please use
%%
\bibliographystyle{elsarticle-num} 
%%  \bibliography{<your bibdatabase>}
%\bibliographystyle{amsplain}
\bibliography{ASIR} 

%% else use the following coding to input the bibitems directly in the
%% TeX file.

%\begin{thebibliography}{00}

%% \bibitem[Author(year)]{label}
%% Text of bibliographic item

%\bibitem[ ()]{}

%\end{thebibliography}
\end{document}